\documentclass[]{JFM-FLM_Au}

\DeclareMathOperator{\f}{f}

\lefttitle{E. Jambon-Puillet}
\righttitle{Journal of Fluid Mechanics}

\title{Hydraulic resistance of channels obstructed by a dense array of elastic fibers}

\author{Etienne Jambon-Puillet\aff{1}}

\affiliation{\aff{1}LadHyX, CNRS, Ecole Polytechnique, Institut Polytechnique de Paris, Palaiseau, France}

\corresau{Etienne Jambon-Puillet, \email{etienne.jambon-puillet@polytechnique.edu}}

\begin{document}
\maketitle

\begin{abstract}
Dense arrays of soft hair-like structures protruding from surfaces are ubiquitous in living systems. Fluid flows can easily deform these soft hairs, which in turn impact the flow properties. At the microscale, flows are often confined which exacerbates this feedback loop: the hair deformation strongly affects the flow geometry. Here, I investigate experimentally and theoretically pressure driven flows in laminar channels obstructed by a dense array of elastic fibers or `hairs'. I show that the system displays a non-linear hydraulic resistance that I model by treating the hair bed as a deformable porous medium whose height results from the deflection of individual fibers. This fluid-structure interaction model encompassing flow in porous media, confinement, and elasticity is then leveraged to identify the key dimensionless parameter governing the problem: $\hat{f}_0$ a dimensionless drag that combines fluid, solid, and geometrical properties. Finally, I demonstrate how these results can be harnessed to design passive flow control elements for microfluidic networks.
\end{abstract}


\section{introduction}

From fur to grass fields, macroscopic surfaces covered with a dense array of slender structures are ubiquitous in our daily lives. When such surfaces are immersed in fluids, the feedback loop between the structure deformation and the flow can alter the flow field around them in non-trivial ways, with potential consequences for living organisms~\citep{deLangre2008,Nepf2012}.
Analogous `hairy' surfaces are also omnipresent at the microscale inside and on the surface of living organisms. Examples include the papilla of tongues~\citep{Harper2013,Lechantre2021}, the brush border in the gut and kidney~\citep{Weinbaum2010}, the cilia in respiratory tracts~\citep{Loiseau2020} or the glycocalyx in blood vessels~\citep{Weinbaum2021}. These hair-like structures perform a wide range of functions that are essential for the survival of many organisms, including humans, such as sensing or filtering for passive ones~\citep{Peter2010,HU2010}, or transporting fluids or cells for active ones~\citep{Gilpin2020}. 

At these small scales, confinement has a strong influence on the flows that are typically laminar and exacerbates the fluid-structure interaction feedback loop. As the structure deforms, it modifies the degree of confinement and thus has a much larger impact on the fluid flow than in unconfined cases where this feedback can often be neglected~\citep{Guglielmini2012,Young2012}. Recently, the interaction between individual elastic structures and confined laminar flows has received a lot of attention since it can be harnessed for passive flow control in microchannels~\citep{Leslie2009,Gomez2017,Louf2020,Brandenbourger2020,vanLaake2022,Moore:2023,Garg2024}. On the other hand, less is known about the fluid-structure interaction of confined dense arrays of elastic structures akin to biological hairy surfaces. There, collective effects come into play and the hairy surface behavior cannot be determined from the study of single hairs~\citep{Alvarado2017,Stein2019,deBlois2023,Ushay2023}. 

While new numerical methods are currently being developed to tackle the fluid-structure interactions in such systems~\citep{Stein2019}, only the `simple shear' flow is so far well understood~\citep{Alvarado2017}; the hairy surface displays a confinement dependent non-linear drag which can be made directional by inclining the hairs. 
Adapting these results to pressure driven flows, arguably more common in biological and microfluidic applications, is not straightforward. The addition of a pressure field fundamentally changes the flow properties. In `simple shear', there is no flow inside the hair bed which simply acts as a deformable boundary for the outer flow. In pressure driven systems, however, a significant flow can develop inside the hair bed 
(see Movie~S1). 
The hair bed itself is permeable and thus part of the fluid domain to be modeled, which usually requires some form of coarse-graining/homogeneisation procedure~\citep{Stein2019}, i.e. extra tools from flows in porous media. 

Here, I investigate experimentally and theoretically pressure driven flows in channels obstructed by a dense array of elastic fibers or `hairs'. I show that the hair flexibility generates a non-linear pressure-flow rate relationship, aka hydraulic resistance, that I rationalize with a reduced order fluid-structure interaction model. 
At the macroscale, the hairy surface is treated as an effective porous layer whose thickness is governed by the deformation of individual hairs at the microscale.
This model is solved numerically to predict the hair deflection and the channel hydraulic resistance without any free parameters. It is then leveraged to identify the key dimensionless number governing the problem: $\hat{f}_0$ a dimensionless drag that combines fluid, solid, and geometrical properties. The system can be considered rigid for $\hat{f}_0 \ll 1$ and significant non-linearities appear when $\hat{f}_0 \gtrsim 1$. These findings can be used to design hairy surfaces for passive flow control in viscous channel networks such as relief valves, flow rectifiers, or memristors. 

\begin{figure}
    \begin{center}
        \includegraphics[width=\columnwidth]{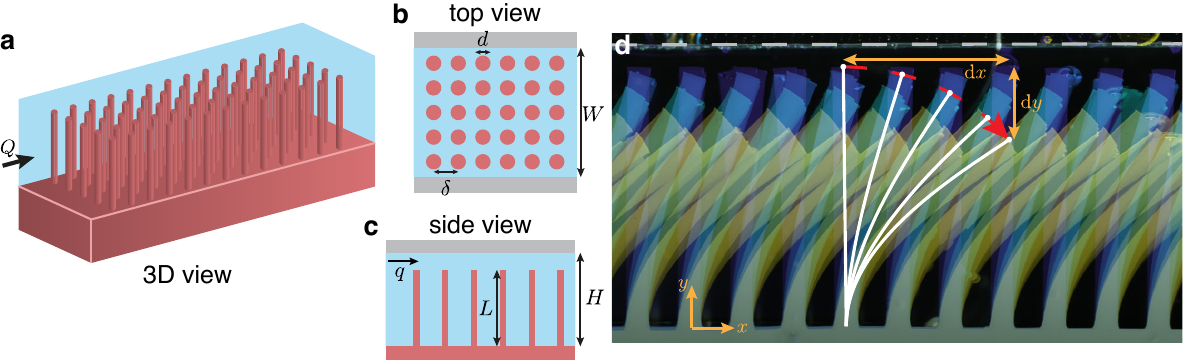}
        \caption{\textbf{(a)} Schematic of the experiment, including cross-sections in the horizontal \textbf{(b)} and vertical \textbf{(c)} directions. The channel walls and ceiling are omitted in (a) for clarity. \textbf{(d)} Superimposition of experimental images of a portion of the channel for different flow rates $Q=\{0,\: 2,\: 5,\: 12,\: 25\}$ mL/min (see also Movie~S1). The coordinate system ($x$,$y$), the centerline of a hair, its tip movement ($\mathrm{d}x$,$\mathrm{d}y$), and the channel ceiling are drawn.
        Parameters: $d= 0.4$ mm, $\delta= 0.83$ mm, $L= 5$ mm, $H\approx 5.5$ mm, $W= 5.4$ mm, $E= 1.27$ MPa, $\eta= 0.89$ Pa.s.
        } 
        \label{fg:fig1}
    \end{center}
\end{figure}

\section{Experimental results}

The experimental setup is sketched in Fig.~\ref{fg:fig1}a-c. Using 3d printed molds, square arrays of elastic hairs of length $L$, diameter $d$, and spacing $\delta$ are molded in silicone elastomers. These hair arrays are then inserted inside a channel in PMMA of width $W$ and height $H$. To avoid swelling the hairs, pure glycerine is used as working fluid. Because glycerine does not wet the silicone elastomer, air gets entrapped inside the hair beds upon filling the channel~\citep{Ushay2023}. To remove the trapped air, the channel is filled without its roof, bubbles are manually removed with a stirrer, and the roof is then installed to fully seal the channel.
The silicone elastomer Young modulus $E$ and the glycerine viscosity $\eta(T)$, with $T$ the room temperature recorded for each experiment, are measured in a rheometer. The fluid flow rate $Q$ is controlled with a syringe pump and the pressures $P_i$ at the inlet and $P_o$ at the outlet of the hair bed are measured with two pressure sensors. The hair deformation is recorded from the side with a camera. More details about the channel manufacturing, material characterization, and experiment procedure are shown in SM. Focusing on the steady-state regime, I explore channels whose free region $H-L\ll W$ such that I expect the problem to be invariant along the channel width and define the reduced 2d flow rate $q=Q/W$. The Reynolds number defined with the largest length scale $H$ is small, $\Rey=\rho q/\eta<0.2$ with $\rho=1260$ kg/m$^3$ the density of glycerine, and the flow is thus laminar.

\begin{figure}
    \begin{center}
        \includegraphics[width=\textwidth]{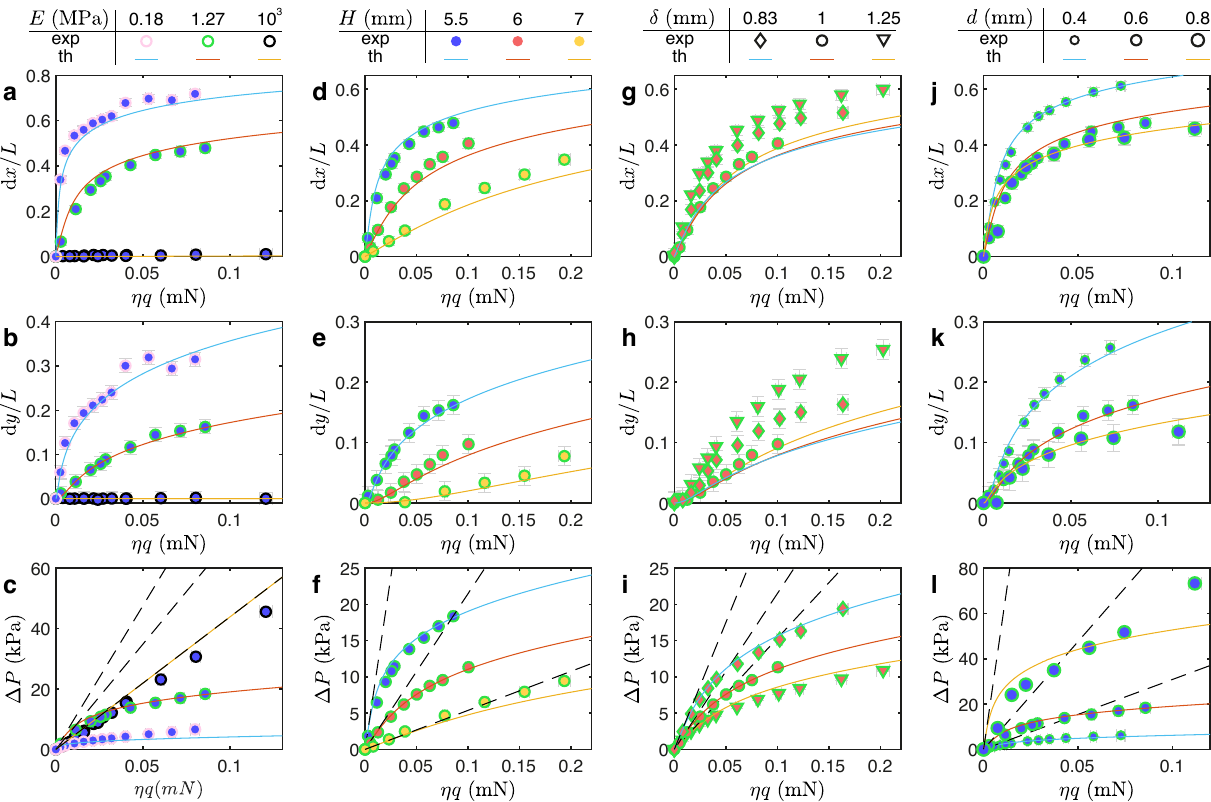}
        \caption{Normalized horizontal and vertical tip deflection $(\mathrm{d}x/L,\mathrm{d}y/L)$, and pressure difference $\Delta P$ as a function of the fluid load $\eta q$ while varying independently the bed and channel parameters. \textbf{(a-c)} Varying the Young modulus $E$ for $H\approx 5.5$ mm, $d=0.6$ mm, $\delta=1$ mm. \textbf{(d-f)} Varying the channel height $H$ for $E=1.27$ MPa, $d=0.6$ mm, $\delta=1$ mm. \textbf{(g-i)} Varying the hair spacing $\delta$ for $E=1.27$ MPa, $H\approx 6$ mm, $d=0.6$ mm. \textbf{(j-l)} Varying the hair diameter $d$ for $E=1.27$ MPa, $H\approx 5.5$ mm, $\delta=1$ mm.
        The parameters are coded through the marker inside ($H$) and outside ($E$) colors, shape ($\delta$), and size ($d$), see legend. Error bars represent the measurement uncertainty. Colored curves are the model prediction, dashed black lines are the pressure-flow rate curves in the absence of deformation.
        } 
        \label{fg:fig2}
    \end{center}
\end{figure}

The side view of the channel during a typical experiment is shown in Fig.~\ref{fg:fig1}d (see also Movie~S1). The hairs, initially straight and obstructing a large portion of the channel, are gradually deformed as the flow rate increases stepwise. 
Fig.~\ref{fg:fig2} shows the normalized horizontal and vertical tip deflection $(\mathrm{d}x/L,\mathrm{d}y/L)$ as well as the pressure difference across the hair bed $\Delta P=P_i-P_o$ as a function of the fluid load $\eta q$ as the hair bed parameters and channel height are varied independently. The fluid load $\eta q$ was chosen as control variable to allow the comparison of experiments done at different room temperatures and thus different viscosities $0.8<\eta (\mathrm{Pa.s}) <1.5$ (see SM).
 
As shown in Fig.~\ref{fg:fig2}a-c, a `rigid' hair bed directly 3d printed with the printer resin barely deforms and exhibits a linear pressure-flow rate relationship. For `soft' hairs, both the horizontal and vertical deflection increase with the fluid load and eventually saturate at very large flow rates. The deformations being large, the hair bed's overall height decreases significantly which opens up the free portion of the channel. As a result, the pressure-flow rate relationship is strongly sub-linear. The initial degree of confinement of the hair bed is varied in Fig.~\ref{fg:fig2}d-f through the channel height $H$. The higher the confinement, the larger the deformations for a given fluid load. A higher confinement increases the overall value of the pressure drop and decreases the fluid load at which it becomes non-linear. Surprisingly, the hair deformation shows small and nonmonotonic variations with the hair spacing $\delta$, as shown in Fig.~\ref{fg:fig2}g-i. However, the pressure drop across the channel does increase monotonicaly as the spacing $\delta$ decreases. Increasing the hair diameter $d$ decreases the hair deformation, since they get stiffer, and increases the pressure drop since it also makes the array tighter (Fig.~\ref{fg:fig2}j-l). 

\section{Fluid-structure interaction model and comparison with experiments}
\subsection{Macroscale fluid velocity}
\label{sec:macro}

\begin{figure}
    \begin{center}
        \includegraphics[width=\columnwidth]{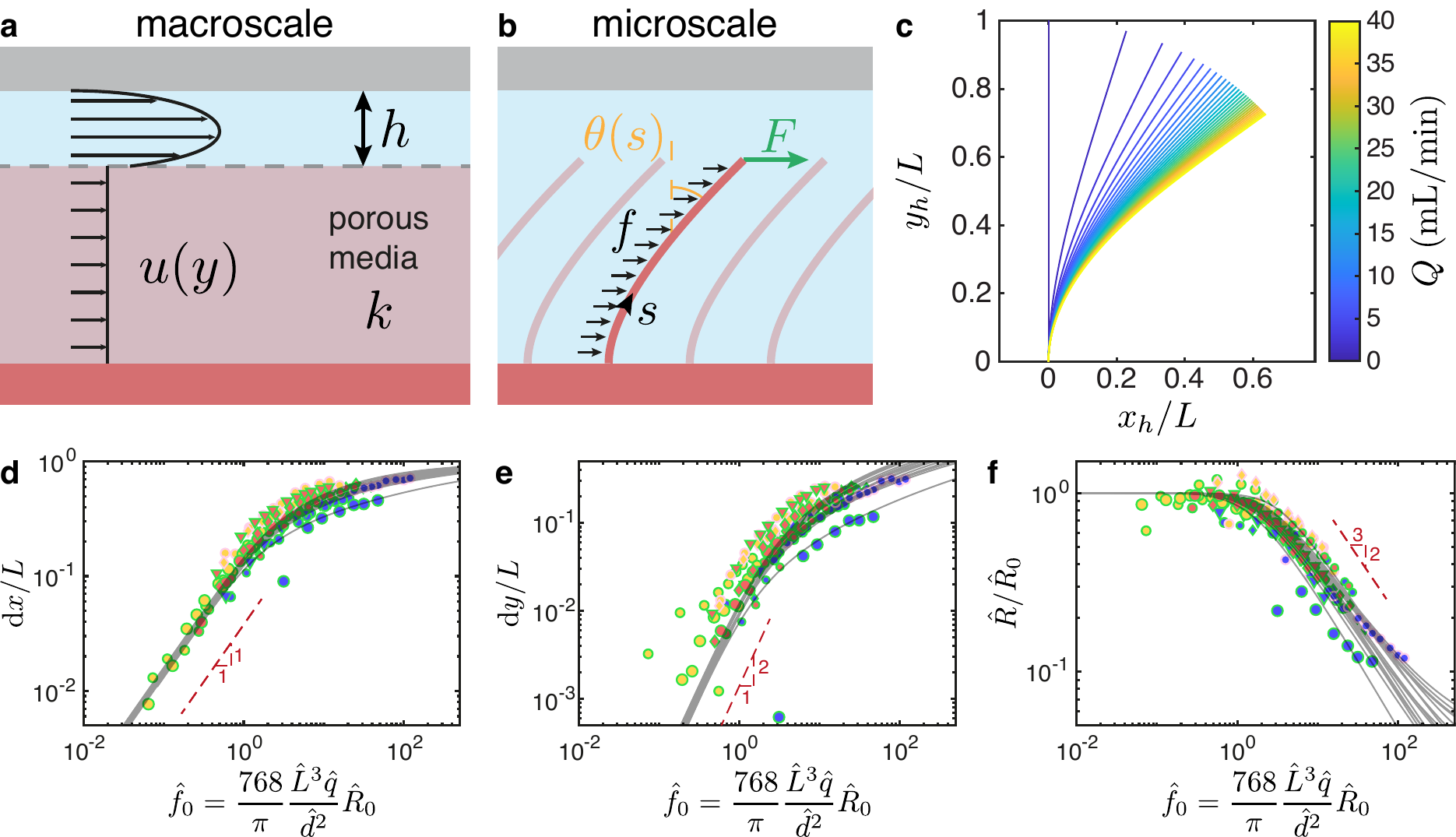}
        \caption{\textbf{(a)} Schematic of flow modeling at the macroscale.
        \textbf{(b)} Schematic of hair mechanics modeling at the microscale.
        \textbf{(c)} Numerical solution of the fluid-structure interaction model, for the experiment in Fig.~\ref{fg:fig1}d. 
        \textbf{(d-f)} Normalized horizontal and vertical tip deflection $(\mathrm{d}x/L,\mathrm{d}y/L)$ (d,e), and normalized hydraulic resistance $\hat{R}/\hat{R}_0$ (f) as a function of the dimensionless drag force $\hat{f}_0$ for all the experiments (see Table~S1). The parameters are coded through the marker inside and outside colors, shape, and size, see legend Fig.~\ref{fg:fig2}. The black curves are the result of the model for the same parameters.
        } 
        \label{fg:fig3}
    \end{center}
\end{figure}

To understand the properties of these hairy channels, I first adopt a macroscale view and model the hair array as an isotropic porous medium of permeability $k$ partially obstructing the channel, as shown in Fig.~\ref{fg:fig3}a. Several approaches exist to model such flows and the conditions at the porous-free interface~\citep{LeBars2006}. For simplicity, I use the \citet{BeaversJoseph1967} formulation that assumes a uniform Darcy flow $u_D=-(k/\eta)\nabla P$ in the porous domain and a Poiseuille flow $u_P$ with a slip length of size $\sqrt{k}/\alpha$ in the free domain of height $h$, with $\alpha$ an empirical coefficient of order $1$ characterizing the porous structure:
\begin{equation}
u_P(y)=-\frac{\nabla P}{2\eta} \left[h^2-(y-(H-h))^2-\frac{\alpha\left(h^2-2k\right)}{\sqrt{k}+\alpha h}\left(H-y\right)\right].
\label{eq:velocity}
\end{equation}
Here $y$ is the vertical direction whose origin is at the base of the hairs (see Fig.~\ref{fg:fig1}d).
The channel hydraulic resistance (per unit width and length) $R=-\nabla P/q$ can then be computed by integrating the flow profile $q=\int_{0}^{H-h} u_D \:\mathrm{d}y+ \int_{H-h}^H u_P \:\mathrm{d}y$:
\begin{equation}
R(h)=\frac{12\eta\left(\sqrt{k}+\alpha h\right)}{\alpha h^4+4\sqrt{k}h^3+6\alpha k h (2H-h)+12(H-h)k^{3/2}}.
\label{eq:ResistHyd}
\end{equation}

Equation~\eqref{eq:ResistHyd} can be readily used to predict the pressure drop for rigid hair beds for which $h=H-L$, provided the permeability $k=\f(d,\delta)$ is known. For regular arrays of cylinders, the permeability can be computed through asymptotic models~\citep{Happel959,Drumonds1984}, scaling laws~\citep{Sobera2006}, or homogenization methods~\citep{Zampogna2016}. For its simplicity and validity over a wide range of packing fractions, I use the scaling $k=c\delta^2\left(1-d/\delta\right)^3$ with $c=0.1475$~\citep{Sobera2006}. Fig.~\ref{fg:fig2}c,f,i,l shows the pressure drop predicted by eq.~\eqref{eq:ResistHyd} with $\alpha=1$ as dashed black lines. It captures reasonably well the pressure drop measured for `rigid' hairs (see Fig.~\ref{fg:fig2}c), as well as the initial slope of the pressure-flow rate relationship of soft hairs. Note that Fig.~\ref{fg:fig2}c shows three curves because the channel height $H$ varies slightly due to the roof installation procedure, resulting in a deviation of $\pm 0.2$ mm  from the target (see measured values in Table~S1).
I also compare the predicted flow profiles, $u_D$ and $u_P$, to simulations of the flow through rigid hair beds in SM and find a reasonable agreement. I therefore set $\alpha=1$ in the following, which is equivalent to setting the slip length at the porous interface to $\sqrt{k}$.

\subsection{Microscale fiber deformation}
To compute the hydraulic resistance of soft hair beds, I need to predict the evolution of the gap $h$ which results from the deformation of individual hairs at the microscale (see Fig.~\ref{fg:fig3}b). Linear beam theory does not allow vertical deformations, and I thus turn to non-linear beam theory. Denoting the coordinates of the hair centerline as $(x_h(s),y_h(s))$, with $s$ the arc-length, and $\theta(s)$ the tangent angle with respect to the vertical, the force and moment balances on a hair read: 
\begin{equation}
	\begin{gathered}
		EI \theta''(s)+\left(F(h)+f(h)(L-s)\right)\cos\theta(s)=0,\\
		y_h'(s)=\cos\theta(s),\qquad x_h'(s)=\sin\theta(s),\\
		x_h(0)=0,\; y_h(0)=0,\; \theta(0)=0,\; \theta'(L)=0.
	\end{gathered}
	\label{eq:beamdim1}
\end{equation}
Here, $(\cdot)'=\mathrm{d}/\mathrm{d}s$, $I=\pi d^4/64$, $f(h)$ is a distributed load coming from the fluid drag inside the hair bed, and $F(h)$ is a concentrated load at the tip from the fluid shear stress in the free region. Both forces a priori depend on the gap $h$ and thus the beam deflection itself since $h=H-y_h(L)=H-\int_0^L \cos(\theta(s))\:\mathrm{d}s$.

To compute these microscale forces, the averaged quantities computed in section \ref{sec:macro} are not adequate and one needs to consider the details of the flow at the hair scale. 
Inside the hair bed, studies on 2d cylinder arrays~\citep{Happel959,Drumonds1984,Sobera2006} have shown that the fluid drag coming from the microscale flow is related to the permeability through momentum conservation: 
\begin{equation}
	f=\frac{\delta^2}{k}\eta u_D=\delta^2 q R(h).
	\label{eq:fdim}
\end{equation}
At the hair tip, the wall shear stress $\tau_s=\eta \left. \frac{\mathrm{d}u}{\mathrm{d}y} \right\vert_{H-h}$ integrated over the tip surface generates the force $F=\int \tau_s \:\mathrm{d}S$. I am not aware of previous studies calculating the local fluid velocity in this configuration. However, it is clear that using the averaged macroscale velocity $u_P$ which is non-zero at the hair surface is inadequate. Guided by the numerical simulations of the flow through rigid hair beds shown in SM, I propose to approximate the local velocity profile as parabolic ${u(y)=\frac{6\bar{u}}{h^2}\left(H-y\right)\left(y-(H-h) \right)}$, with $\bar{u}$ the local average flow velocity. I then use the macro model to estimate this average local velocity $\bar{u}=\frac{1}{h}\int_{H-h}^H u_P \:\mathrm{d}y$. Using this approximate local velocity profile to compute the shear stress and integrating it over the hair tip yields the tip force 
\begin{equation}
	F=\frac{\pi d^2}{4}\eta \left. \frac{\mathrm{d}u}{\mathrm{d}y} \right\vert_{H-h}=\frac{\pi}{8} d^2 q R(h) \frac{\alpha h^2+4\sqrt{k}h+6\alpha k}{\alpha h +\sqrt{k}}.
	\label{eq:Fdim}
\end{equation}

\subsection{Non-dimensionalisation}
Combining equations \eqref{eq:ResistHyd}-\eqref{eq:Fdim} yields a reduced order, fully-coupled, fluid-structure interaction model that has the form of a 1d integro-differential boundary value problem. 
This model is made dimensionless by rescaling all lengths by $L$ in the beam equation~\eqref{eq:beamdim1}
\begin{equation}
	\begin{gathered}
		\theta''(\hat{s})+\left(\hat{F}+\hat{f}(1-\hat{s})\right)\cos\theta(\hat{s})=0,\\
		\hat{y_h}'(\hat{s})=\cos\theta(\hat{s}),\qquad \hat{x_h}'(\hat{s})=\sin\theta(\hat{s}),\\
		\hat{x_h}(0)=0,\; \hat{y_h}(0)=0,\; \theta(0)=0,\; \theta'(1)=0.
	\end{gathered}
	\label{eq:beamadim1}
\end{equation}
Here $\hat{s}=s/L$, $\hat{x_h}=x_h/L$, $\hat{y_h}=y_h/L$, and the dimensionless fluid forces are $\hat{F}=FL^2/EI$, $\hat{f}=fL^3/EI$. Using the relevant aspect ratios, $\hat{L}=L/H$, $\hat{d}=d/\delta$, $\hat{\delta}=\delta/H$, the dimensionless permeability reads $\hat{k}=k/H^2=c\hat{\delta}^2(1-\hat{d})^{3/4}$ and the dimensionless gap ${\hat{h}=h/H=1-\hat{L}\int_0^1\cos\theta\:\mathrm{d}\hat{s}=1-\hat{L}\hat{y_h}(1)}$. Using these along the elasto-viscous parameter $\hat{q}=\eta q/E d^2$ yields the dimensionless forces
\begin{equation}
		\hat{f}=\frac{768}{\pi}\frac{\hat{L}^3\hat{q}}{\hat{d}^2}\hat{R}, \quad
		\hat{F}=96\hat{L}^2\hat{q}\hat{R}\frac{\alpha(1-\hat{L}\hat{y_h}(1))^2+4\sqrt{\hat{k}}(1-\hat{L}\hat{y_h}(1))+6\alpha\hat{k}}{\alpha(1-\hat{L}\hat{y_h}(1))+\sqrt{\hat{k}}},\\
	\label{eq:forceadim}
\end{equation}
with $\hat{R}=RH^3/12\eta$ the hydraulic resistance rescaled by that of an empty channel: 
\begin{equation}
\hat{R}=\frac{\sqrt{\hat{k}}+\alpha (1-\hat{L}\hat{y_h}(1))}{\alpha (1-\hat{L}\hat{y_h}(1))^4+4\sqrt{\hat{k}}(1-\hat{L}\hat{y_h}(1))^3+6\alpha \hat{k} (1-(\hat{L}\hat{y_h}(1))^2)+12\hat{L}\hat{y_h}(1)\hat{k}^{3/2}}.
\label{eq:resistHydadim}
\end{equation}

I solve the dimensionless equations \eqref{eq:beamadim1}\eqref{eq:forceadim}\eqref{eq:resistHydadim} numerically with the Matlab solver bvp5c and show in Fig.~\ref{fg:fig3}c the hair profile as the flow rate is increased for the experimental parameters of Fig.~\ref{fg:fig1}d. In Fig.~\ref{fg:fig2}, I compare the tip deflection predicted by the model to the one measured experimentally. I find a quantitative agreement for most experimental conditions without any adjustable parameters. In panels (g,h), the permeability spans an order of magnitude, and the quantitative discrepancies in the model are likely due to inaccuracies in the scaling used to calculate $k$. Despite these limits, the model captures the surprisingly weak influence of the hair spacing $\delta$ on the hair deflection seen in these experiments. Using the computed tip deflection for $h$ in eq.~\eqref{eq:ResistHyd} yields a prediction for the pressure drop across the channel which is compared to the experimental pressure drop in Fig.~\ref{fg:fig2}c,f,i,l. Here also the agreement is excellent. 

\subsection{Asymptotic expansion for small flow rates}
To get analytical insights on the role of the various length scales in the problem, I first notice that the tip force $\hat{F}$ is usually negligible compared to the drag inside the hair bed $\hat{f}$:
\[\frac{\hat{f}}{\hat{F}}=\frac{8}{\pi}\frac{\hat{L}}{\hat{d}^2}\frac{\alpha(1-\hat{L}\hat{y_h}(1))+\sqrt{\hat{k}}}{\alpha(1-\hat{L}\hat{y_h}(1))^2+4\sqrt{\hat{k}}(1-\hat{L}\hat{y_h}(1))+6\alpha\hat{k}}\gg 1.\]
Since $\hat{q}<2.3\cdot 10^{-3}$ in experiments, I then introduce the asymptotic expansion ${\theta(\hat{s})=\sum_n \hat{q}^n \theta_n(\hat{s})}$, with $\hat{q}$ treated as a small parameter~\citep{Alvarado2017} in eq.~\eqref{eq:beamadim1} with $\hat{F}=0$. Restricting myself to the case of straight hairs for simplicity, the solution should be symmetric with respect to the flow direction and even terms are thus zero. In particular, at order zero $\theta_0(\hat{s})=0$. At the first order in $\hat{q}$, $\theta(\hat{s})= \hat{q} \theta_1(\hat{s})$ and thus $\cos\theta(\hat{s})=\cos(\hat{q} \theta_1(\hat{s}))= 1$, and $\hat{y_h}(1)=\int_0^1\cos\theta(\hat{s})\:\mathrm{d}\hat{s}= 1$. Inserting this in the beam equation without tip force yields at first order
\begin{equation}
\hat{q} \theta_1''(\hat{s})+\hat{f}_0(1-\hat{s})=0,
\label{eq:asympt1}
\end{equation}
with 
\begin{equation}
\hat{f}_0=\frac{768}{\pi}\frac{\hat{L}^3\hat{q}}{\hat{d}^2}\hat{R}_0,\quad
\hat{R}_{0}=\frac{\sqrt{\hat{k}}+\alpha(1-\hat{L})}{\alpha (1-\hat{L})^4+4\sqrt{\hat{k}}(1-\hat{L})^3+6\alpha  \hat{k} (1-\hat{L}^2)+12\hat{L}\hat{k}^\frac{3}{2}}.
	\label{eq:forcelin}
\end{equation}
Here $\hat{R}_0$ is the dimensionless hydraulic resistance in the absence of deformations, i.e. for $\hat{y_h}(1)=1$. Eq.~\eqref{eq:asympt1} is mathematically equivalent to a linear beam model under a uniform distributed load whose solution is 
\begin{equation}
\theta_1(\hat{s})=\frac{\hat{f}_0}{6 \hat{q}}\left(\hat{s}^3-3\hat{s}^2+\hat{s}\right).
\label{eq:asymptsol}
\end{equation}
Since $\theta_2(\hat{s})=0$ by symmetry, $\theta(\hat{s})= \hat{q} \theta_1(\hat{s})$ with $\theta_1(\hat{s})$ from eq.~\eqref{eq:asymptsol} is valid up to second order in $\hat{q}$. 

This analysis suggests that the four dimensionless parameters $\hat{q}$, $\hat{L}$, $\hat{\delta}$, and $\hat{d}$ can be combined in a single one $\hat{f}_0$ (see eq.~\eqref{eq:forcelin}). This small $\hat{q}$ / small slope solution can then be used to compute the tip deflection
\begin{equation}
	\begin{gathered}
	\hat{x_h}(1)=\int_0^1 \sin\theta \:\mathrm{d}\hat{s}\approx \int_0^1  \hat{q} \theta_1 \:\mathrm{d}\hat{s}\approx \frac{\hat{f}_0}{8},\\
	\hat{y_h}(1)=\int_0^1 \cos\theta \:\mathrm{d}\hat{s}\approx \int_0^1 \left(1-\frac{(\hat{q} \theta_1)^2}{2}\right) \:\mathrm{d}\hat{s}\approx 1- \frac{\hat{f}^2_0}{112}.
		\end{gathered}
	\label{eq:solsmalparamtip}
\end{equation}

Fig.~\ref{fg:fig3}d,e shows the tip deflection for all the experiments and the model plotted as a function of $\hat{f}_0$. Surprisingly, the data not only collapse for small deformations, as expected but also for larger deformations where the asymptotic expansion is expected to fail. The hydraulic resistance $\hat{R}$, computed from the measured pressure loss $\Delta P$ for experiments and calculated from the tip deflection using eq.~\eqref{eq:resistHydadim} for the model, is rescaled by its value for rigid hairs $\hat{R}_0$ and shown as a function of $\hat{f}_0$ in Fig.~\ref{fg:fig3}f. Here also the data collapse, even for large deflection. The hair flexibility starts to impact the hydraulic resistance when $\hat{f}_0\sim 1$, and I observe empirically $\hat{R}/\hat{R}_{0}\sim \hat{f}^{-2/3}_0$ for large deformations. The rigid hair bed drag force $\hat{f}_0$ is thus an efficient control parameter encompassing both the material and geometrical properties of the system and can be used to predict the hydraulic resistance of channels obstructed by dense hair-like structures. 

\section{Discussion and conclusion}
\subsection{Model assumptions, limitations and alternative approaches}
The fluid-structure interaction model developed above prioritizes simplicity in order to be as easy to interpret as possible while still reproducing the experiments. To achieve this, several simplifying assumptions were made, which introduces limitations. These limitations, as well as alternative modeling approaches to overcome them are discussed below.

On the fluid side, the \citet{BeaversJoseph1967} formulation assumes that we can separate a porous region from a free region. Moreover, the Darcy flow used in the porous domain neglects the boundary layers of size $\sim\sqrt{k}$ that develop at its edges (see SM). While the matching to the top boundary layer is accounted for in the free region via the slip length, both are neglected in the porous region. These boundary layers most notably impact the drag force $f$ that is no longer constant along the arclength. Both of these assumptions are reasonable when the typical pore size is the smallest length scale in the problem, i.e. $L\gg \sqrt{k}$ and $h\gg \sqrt{k}$. This is mostly the case case in the present experiments (see Table~S1). For sparser hair beds, a more complex Darcy-Brinkman approach~\citep{LeBars2006,Zampogna2016} could be used to overcome both of these limitations. Since the Darcy-Brinkman flow field $u(y)$ has an analytical solution in 2d, its implementation in the beam equation should be straightforward. 
Finally, the model assumes invariance along the channel width and is thus not applicable to narrow channels ($H-L>W$). In that case, a Hele-Shaw approach could be more suitable to treat the flow in the free region~\citep{Wexler2013}. 

On the solid side, the beam model does not account for hair-to-hair contact which would effectively `rigidify' the hairs. If the hair bed is tight enough to be approximated as a porous medium, then $L>\delta$ and contact is bound to happen at sufficiently large deformations. The critical contact drag force $\hat{f}_0$ depends on how tight the packing is, i.e. $\delta-d$, and can be computed from the present model by checking whether two computed beam profiles of thickness $d$ separated by $\delta$ intersect. To study this contact regime, one could simulate several beams and implement contact forces in the beam model, which is more conveniently done on discrete models \citep{Bergou:2008}.

On the porous medium side, the change in permeability due to the hair deformation is not accounted for. In the Cartesian reference frame of the lab, as the hairs deform, their center to center distance changes in the ($x$) direction and their ($x$,$z$) cross section becomes elliptical. Alternatively, we can adopt the fiber local reference frame. While the fiber cross section remains circular, the flow direction is no longer perpendicular to the fiber. In both cases, accounting for the hair deformation is not trivial and would yield a deformation dependent anisotropic permeability $k(y,\theta(s))$.

\begin{figure}
    \begin{center}
        \includegraphics[width=0.7\columnwidth]{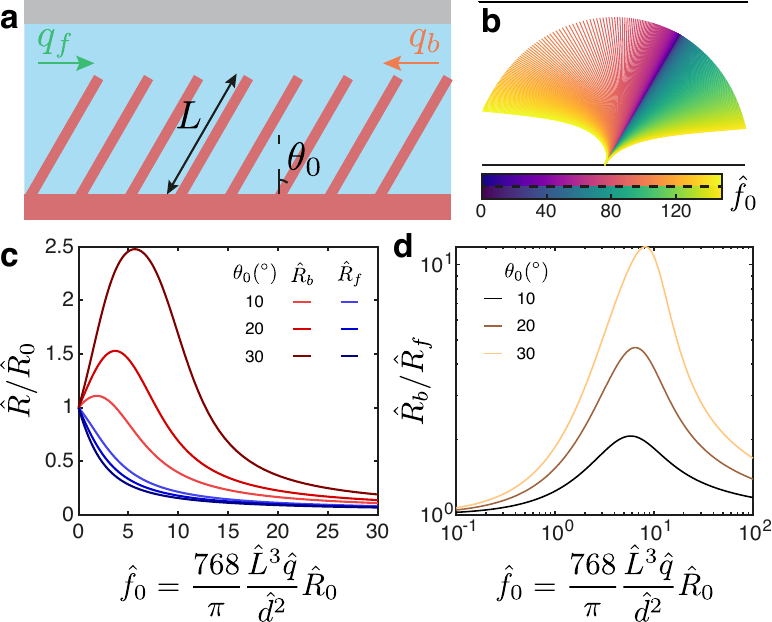}
        \caption{\textbf{(a)} Schematic of a channel with hairs inclined at an angle $\theta_0$. The channel is now asymmetric with respect to the flow direction, forward $q_f$ or backward $q_b$. 
        \textbf{(b)} Model prediction for the deformation of inclined hairs with $\theta_0=30^\circ$. Other parameters are similar to Fig.~\ref{fg:fig1}d ($\hat{L}\approx 0.92$, $\hat{d}=0.5$, $\hat{\delta}=0.15$). The black lines indicate the channel walls and the two color maps indicate the dimensionless drag force $\hat{f}_0$ in the forward and backward directions.
        \textbf{(c)} Predicted normalized hydraulic resistance $\hat{R}/\hat{R}_0$ in the forward and backward direction as a function of $\hat{f}_0$ for three inclination angles ($\hat{d}=0.5$, $\hat{\delta}=0.15$). The initial effective porous media height, and thus $R_0$, are kept constant by setting $\hat{L}=0.8/\cos\theta_0$.  
        \textbf{(d)} Ratio of the backward to forward resistance $\hat{R}_b/\hat{R}_f$ shown in c.
        } 
        \label{fg:fig4}
    \end{center}
\end{figure}

Finally, a more general numerical approach such as the one proposed by \citet{Stein2019} could also be used. Although computationally more expensive, it can be used on a wider range of geometries.

\subsection{Potential for applications}
The highly non-linear pressure-flow rate relationship of channels obstructed by a dense array of soft hairs has several applications for passive flow control in microfluidic networks. Straight hairs can be used as a relief valve that triggers when $\hat{f}_0> 1$. At normal operating pressure ($\hat{f}_0\ll 1$), the flow rate in a dense and confined hairy channel remains low but if there is any pressure build-up above $\hat{f}_0> 1$ the channel opens up and releases the pressure quickly. Breaking the system symmetry by inclining the hairs (see Fig.~\ref{fg:fig4}a) induces a direction dependent hydraulic resistance~\citep{Alvarado2017,Stein2019}. When the flow is against the grain, the hydraulic resistance increases instead of decreasing which could be used as a fuse~\citep{Box2020} or flow rectifier~\citep{Alvarado2017,Stein2019}. Assembling several of these in networks could act as a memristor and encode information as recently demonstrated with soft valve networks~\citep{Martinez-Calvo2024}. 

For all these applications, confinement is key. While the model remains valid when the hair bed does not fill the majority of the channel, the non-linear effects weaken drastically. Both the dimensionless drag force $\hat{f}_0$ and the ratio  $\hat{R}_0$ between the undeformed and fully deformed hydraulic resistance decrease rapidly with $\hat{L}$. Nonetheless, the reduced order fluid-structure interaction model presented here, which is solved almost instantly on a laptop, can be leveraged to guide the design of hair beds for applications. 

As a proof of concept, I look at the flow rectification application with inclined hairs by simply changing the boundary condition $\theta(0)=\theta_0$ in eq.~\eqref{eq:beamadim1}. Numerical beam profiles for a hair bed similar to the one of Fig.~\ref{fg:fig1}d but inclined by $\theta_0=30^\circ$ are shown in Fig.~\ref{fg:fig4}b for forward and backward flows with different colormaps. In Fig.~\ref{fg:fig4}c I show the hydraulic resistance in the forward $\hat{R}_f$ and backward $\hat{R}_b$ direction as a function of the dimensionless drag force $\hat{f}_0$ for the same hair bed at three levels of inclination (the effective porous media height $\hat{L}\cos\theta_0$ is kept constant). The backward hydraulic resistance passes through a maximum for $\hat{f}_0\sim 5-10$ signaling that there is an optimal flow rate for rectification that is evidenced in Fig.~\ref{fg:fig4}d which plots $\hat{R}_b/\hat{R}_f$ as a function of $\hat{f}_0$.

\subsection{Conclusion}
In summary, fluidic channels obstructed by a dense array of elastic hairs display a highly non-linear pressure-flow rate relationship that results from hair deformations. This non-linearity is rationalized with a macroscale two region flow model where the hair bed is treated as a deformable porous medium whose height is governed by the deflection of individual fibers at the microscale. This reduced order, fully coupled, fluid-structure interaction model is solved numerically and shows a good agreement with experiments, without any fitting parameter. An asymptotic analysis allows to extract the key dimensionless parameter governing the hair response $\hat{f}_0$ that combines an elasto-viscous parameter with the several length scales of the problem. Finally, I show how to leverage the model to design passive flow control elements in fluidic networks based on these hairy channels.

\begin{bmhead}[Supplementary material.]
See Supplementary Materials at [URL will be inserted by publisher] for experimental methods, material characterization, numerical simulations of flow rigid hair beds, a table summarizing the parameters of each experiment, and a supplementary movie of the experiment.
\end{bmhead}

\begin{bmhead}[Acknowledgments.]
I thank A. Rocchi for his help with the mold design and the Drahi-X Novation Center for the use of their equipment.
\end{bmhead}

\begin{bmhead}[Declaration of interests.]
The author report no conflict of interest.
\end{bmhead}

\begin{bmhead}[Funding.]
This research was funded, in whole or in part, by l'Agence Nationale de la Recherche (ANR), project ANR-24-CE30-5885. For the purpose of open access, the author has applied a CC BY public copyright license to any Author Accepted Manuscript (AAM) version arising from this submission.
\end{bmhead}

\bibliographystyle{jfm}
\bibliography{Poils_bib}

@article{BeaversJoseph1967, 
title={Boundary conditions at a naturally permeable wall}, 
volume={30}, 
DOI={10.1017/S0022112067001375}, 
number={1}, 
journal={Journal of Fluid Mechanics}, 
author={Beavers, Gordon S. and Joseph, Daniel D.}, 
year={1967}, 
pages={197–207}
}

@Article{Alvarado2017,
author={Alvarado, Jos{\'e} and Comtet, Jean and de Langre, Emmanuel and Hosoi, A. E.},
title={Nonlinear flow response of soft hair beds},
journal={Nature Physics},
year={2017},
month={Oct},
day={01},
volume={13},
number={10},
pages={1014-1019},
issn={1745-2481},
doi={10.1038/nphys4225},
url={https://doi.org/10.1038/nphys4225}
}

@article{LeBars2006, 
title={Interfacial conditions between a pure fluid and a porous medium: implications for binary alloy solidification}, 
volume={550}, 
DOI={10.1017/S0022112005007998}, 
journal={Journal of Fluid Mechanics}, 
author={Le Bars, Michael and Worster, Michael Grae}, 
year={2006}, 
pages={149–173}
}

@article{Drumonds1984,
title = {Laminar viscous flow through regular arrays of parallel solid cylinders},
journal = {International Journal of Multiphase Flow},
volume = {10},
number = {5},
pages = {515-540},
year = {1984},
issn = {0301-9322},
doi = {https://doi.org/10.1016/0301-9322(84)90079-X},
url = {https://www.sciencedirect.com/science/article/pii/030193228490079X},
author = {J.E. Drummond and M.I. Tahir}
}

@article{Sobera2006,
  title = {Hydraulic permeability of ordered and disordered single-layer arrays of cylinders},
  author = {Sobera, M. P. and Kleijn, C. R.},
  journal = {Phys. Rev. E},
  volume = {74},
  issue = {3},
  pages = {036301},
  numpages = {10},
  year = {2006},
  month = {Sep},
  publisher = {American Physical Society},
  doi = {10.1103/PhysRevE.74.036301},
  url = {https://link.aps.org/doi/10.1103/PhysRevE.74.036301}
}

@article{Nepf2012,
   author = "Nepf, Heidi M.",
   title = "Flow and Transport in Regions with Aquatic Vegetation", 
   journal= "Annual Review of Fluid Mechanics",
   year = "2012",
   volume = "44",
   number = "Volume 44, 2012",
   pages = "123-142",
   doi = "https://doi.org/10.1146/annurev-fluid-120710-101048",
   url = "https://www.annualreviews.org/content/journals/10.1146/annurev-fluid-120710-101048",
   publisher = "Annual Reviews",
   issn = "1545-4479",
   type = "Journal Article",
  }

@article{deLangre2008,
   author = "de Langre, Emmanuel",
   title = "Effects of Wind on Plants", 
   journal= "Annual Review of Fluid Mechanics",
   year = "2008",
   volume = "40",
   number = "Volume 40, 2008",
   pages = "141-168",
   doi = "https://doi.org/10.1146/annurev.fluid.40.111406.102135",
   url = "https://www.annualreviews.org/content/journals/10.1146/annurev.fluid.40.111406.102135",
   publisher = "Annual Reviews",
   issn = "1545-4479",
   type = "Journal Article",
  }

@article{Ushay2023,
  title = {Interfacial flows past arrays of elastic fibers},
  author = {Ushay, C. and Jambon-Puillet, E. and Brun, P.-T.},
  journal = {Phys. Rev. Fluids},
  volume = {8},
  issue = {4},
  pages = {044001},
  numpages = {10},
  year = {2023},
  month = {Apr},
  publisher = {American Physical Society},
  doi = {10.1103/PhysRevFluids.8.044001},
  url = {https://link.aps.org/doi/10.1103/PhysRevFluids.8.044001}
}

@Article{Weinbaum2021,
author={Weinbaum, Sheldon
and Cancel, Limary M.
and Fu, Bingmei M.
and Tarbell, John M.},
title={The Glycocalyx and Its Role in Vascular Physiology and Vascular Related Diseases},
journal={Cardiovascular Engineering and Technology},
year={2021},
month={Feb},
day={01},
volume={12},
number={1},
pages={37-71},
issn={1869-4098},
doi={10.1007/s13239-020-00485-9},
url={https://doi.org/10.1007/s13239-020-00485-9}
}

@article{Harper2013,
author = {Cally J. Harper  and Sharon M. Swartz  and Elizabeth L. Brainerd },
title = {Specialized bat tongue is a hemodynamic nectar mop},
journal = {Proceedings of the National Academy of Sciences},
volume = {110},
number = {22},
pages = {8852-8857},
year = {2013},
doi = {10.1073/pnas.1222726110},
URL = {https://www.pnas.org/doi/abs/10.1073/pnas.1222726110},
OPTeprint = {https://www.pnas.org/doi/pdf/10.1073/pnas.1222726110}}

@article{Peter2010,
    author = {Satir, Peter and Pedersen, Lotte B. and Christensen, Søren T.},
    title = {The primary cilium at a glance},
    journal = {Journal of Cell Science},
    volume = {123},
    number = {4},
    pages = {499-503},
    year = {2010},
    month = {02},
    issn = {0021-9533},
    doi = {10.1242/jcs.050377},
    url = {https://doi.org/10.1242/jcs.050377},
    OPTeprint = {https://journals.biologists.com/jcs/article-pdf/123/4/499/1452641/499.pdf},
}

@Article{Gilpin2020,
author={Gilpin, William
and Bull, Matthew Storm
and Prakash, Manu},
title={The multiscale physics of cilia and flagella},
journal={Nature Reviews Physics},
year={2020},
month={Feb},
day={01},
volume={2},
number={2},
pages={74-88},
issn={2522-5820},
doi={10.1038/s42254-019-0129-0},
url={https://doi.org/10.1038/s42254-019-0129-0}
}

@Article{Loiseau2020,
author={Loiseau, Etienne
and Gsell, Simon
and Nommick, Aude
and Jomard, Charline
and Gras, Delphine
and Chanez, Pascal
and D'Ortona, Umberto
and Kodjabachian, Laurent
and Favier, Julien
and Viallat, Annie},
title={Active mucus--cilia hydrodynamic coupling drives self-organization of human bronchial epithelium},
journal={Nature Physics},
year={2020},
month={Nov},
day={01},
volume={16},
number={11},
pages={1158-1164},
issn={1745-2481},
doi={10.1038/s41567-020-0980-z},
url={https://doi.org/10.1038/s41567-020-0980-z}
}

@article{Weinbaum2010,
author = {Weinbaum, Sheldon and Duan, Yi and Satlin, Lisa M. and Wang, Tong and Weinstein, Alan M.},
title = {Mechanotransduction in the renal tubule},
journal = {American Journal of Physiology-Renal Physiology},
volume = {299},
number = {6},
pages = {F1220-F1236},
year = {2010},
doi = {10.1152/ajprenal.00453.2010},
note ={PMID: 20810611},
URL = {https://doi.org/10.1152/ajprenal.00453.2010},
OPTeprint = {https://doi.org/10.1152/ajprenal.00453.2010}
}

@Article{HU2010,
author={Riisg{\aa}rd, Hans Ulrik and Larsen, Poul S},
title={Particle capture mechanisms in suspension-feeding invertebrates},
journal={Marine Ecology Progress Series},
year={2010},
volume={418},
pages={255-293},
doi={10.3354/meps08755},
url={https://www.int-res.com/abstracts/meps/v418/p255-293/},
url={https://doi.org/10.3354/meps08755}
}

@article{Gomez2017,
  title = {Passive Control of Viscous Flow via Elastic Snap-Through},
  author = {Gomez, Michael and Moulton, Derek E. and Vella, Dominic},
  journal = {Phys. Rev. Lett.},
  volume = {119},
  issue = {14},
  pages = {144502},
  numpages = {5},
  year = {2017},
  month = {Oct},
  publisher = {American Physical Society},
  doi = {10.1103/PhysRevLett.119.144502},
  url = {https://link.aps.org/doi/10.1103/PhysRevLett.119.144502}
}

@article{Brandenbourger2020,
  title = {Tunable flow asymmetry and flow rectification with bio-inspired soft leaflets},
  author = {Brandenbourger, M. and Dangremont, A. and Sprik, R. and Coulais, C.},
  journal = {Phys. Rev. Fluids},
  volume = {5},
  issue = {8},
  pages = {084102},
  numpages = {11},
  year = {2020},
  month = {Aug},
  publisher = {American Physical Society},
  doi = {10.1103/PhysRevFluids.5.084102},
  url = {https://link.aps.org/doi/10.1103/PhysRevFluids.5.084102}
}

@Article{vanLaake2022,
author={van Laake, Lucas C.
and de Vries, Jelle
and Malek Kani, Sevda
and Overvelde, Johannes T.B.},
title={A fluidic relaxation oscillator for reprogrammable sequential actuation in soft robots},
journal={Matter},
year={2022},
month={Sep},
day={07},
publisher={Elsevier},
volume={5},
number={9},
pages={2898-2917},
issn={2590-2393},
doi={10.1016/j.matt.2022.06.002},
url={https://doi.org/10.1016/j.matt.2022.06.002}
}

@article{Wexler2013, title={Bending of elastic fibres in viscous flows: the influence of confinement}, volume={720}, DOI={10.1017/jfm.2013.49}, journal={Journal of Fluid Mechanics}, author={Wexler, Jason S. and Trinh, Philippe H. and Berthet, Helene and Quennouz, Nawal and du Roure, Olivia and Huppert, Herbert E. and Lindner, Anke and Stone, Howard A.}, year={2013}, pages={517–544}}

@article{Garg2024,
  title = {Passive Viscous Flow Selection via Fluid-Induced Buckling},
  author = {Garg, Hemanshul and Ledda, Pier Giuseppe and Pedersen, Jon Skov and Pezzulla, Matteo},
  journal = {Phys. Rev. Lett.},
  volume = {133},
  issue = {8},
  pages = {084001},
  numpages = {5},
  year = {2024},
  month = {Aug},
  publisher = {American Physical Society},
  doi = {10.1103/PhysRevLett.133.084001},
  url = {https://link.aps.org/doi/10.1103/PhysRevLett.133.084001}
}

@article{deBlois2023,
  title = {Canopy elastic turbulence: Spontaneous formation of waves in beds of slender microposts},
  author = {de Blois, Charlotte and Haward, Simon J. and Shen, Amy Q.},
  journal = {Phys. Rev. Fluids},
  volume = {8},
  issue = {2},
  pages = {023301},
  numpages = {21},
  year = {2023},
  month = {Feb},
  publisher = {American Physical Society},
  doi = {10.1103/PhysRevFluids.8.023301},
  url = {https://link.aps.org/doi/10.1103/PhysRevFluids.8.023301}
}

@article{Stein2019,
  title = {Coarse graining the dynamics of immersed and driven fiber assemblies},
  author = {Stein, David B. and Shelley, Michael J.},
  journal = {Phys. Rev. Fluids},
  volume = {4},
  issue = {7},
  pages = {073302},
  numpages = {29},
  year = {2019},
  month = {Jul},
  publisher = {American Physical Society},
  doi = {10.1103/PhysRevFluids.4.073302},
  url = {https://link.aps.org/doi/10.1103/PhysRevFluids.4.073302}
}

@article{Guglielmini2012,
    author = {Guglielmini, Laura and Kushwaha, Amit and Shaqfeh, Eric S. G. and Stone, Howard A.},
    title = {Buckling transitions of an elastic filament in a viscous stagnation point flow},
    journal = {Physics of Fluids},
    volume = {24},
    number = {12},
    pages = {123601},
    year = {2012},
    month = {12},
    issn = {1070-6631},
    doi = {10.1063/1.4771606},
    url = {https://doi.org/10.1063/1.4771606},
    OPTeprint = {https://pubs.aip.org/aip/pof/article-pdf/doi/10.1063/1.4771606/13433094/123601\_1\_online.pdf},
}

@Article{Young2012,
author={Young, Y.-N.
and Downs, M.
and Jacobs, C. R.},
title={Dynamics of the Primary Cilium in Shear Flow},
journal={Biophysical Journal},
year={2012},
month={Aug},
day={22},
publisher={Elsevier},
volume={103},
number={4},
pages={629-639},
issn={0006-3495},
doi={10.1016/j.bpj.2012.07.009},
url={https://doi.org/10.1016/j.bpj.2012.07.009}
}

@article{Louf2020,
  title = {Bending and Stretching of Soft Pores Enable Passive Control of Fluid Flows},
  author = {Louf, Jean-Fran\ifmmode \mbox{\c{c}}\else \c{c}\fi{}ois and Knoblauch, Jan and Jensen, Kaare H.},
  journal = {Phys. Rev. Lett.},
  volume = {125},
  issue = {9},
  pages = {098101},
  numpages = {6},
  year = {2020},
  month = {Aug},
  publisher = {American Physical Society},
  doi = {10.1103/PhysRevLett.125.098101},
  url = {https://link.aps.org/doi/10.1103/PhysRevLett.125.098101}
}

@article{Lechantre2021,
author = {Amandine Lechantre  and Ayrton Draux  and Hoa-Ai Béatrice Hua  and Denis Michez  and Pascal Damman  and Fabian Brau },
title = {Essential role of papillae flexibility in nectar capture by bees},
journal = {Proceedings of the National Academy of Sciences},
volume = {118},
number = {19},
pages = {e2025513118},
year = {2021},
doi = {10.1073/pnas.2025513118},
URL = {https://www.pnas.org/doi/abs/10.1073/pnas.2025513118},
OPTeprint = {https://www.pnas.org/doi/pdf/10.1073/pnas.2025513118}}

@Article{Leslie2009,
author={Leslie, Daniel C.
and Easley, Christopher J.
and Seker, Erkin
and Karlinsey, James M.
and Utz, Marcel
and Begley, Matthew R.
and Landers, James P.},
title={Frequency-specific flow control in microfluidic circuits with passive elastomeric features},
journal={Nature Physics},
year={2009},
month={Mar},
day={01},
volume={5},
number={3},
pages={231-235},
issn={1745-2481},
doi={10.1038/nphys1196},
url={https://doi.org/10.1038/nphys1196}
}

@Article{Martinez-Calvo2024,
author={Mart{\'i}nez-Calvo, Alejandro
and Biviano, Matthew D.
and Christensen, Anneline H.
and Katifori, Eleni
and Jensen, Kaare H.
and Ruiz-Garc{\'i}a, Miguel},
title={The fluidic memristor as a collective phenomenon in elastohydrodynamic networks},
journal={Nature Communications},
year={2024},
month={Apr},
day={10},
volume={15},
number={1},
pages={3121},
issn={2041-1723},
doi={10.1038/s41467-024-47110-0},
url={https://doi.org/10.1038/s41467-024-47110-0}
}

@article{Happel959,
author = {Happel, John},
title = {Viscous flow relative to arrays of cylinders},
journal = {AIChE Journal},
volume = {5},
number = {2},
pages = {174-177},
doi = {https://doi.org/10.1002/aic.690050211},
url = {https://aiche.onlinelibrary.wiley.com/doi/abs/10.1002/aic.690050211},
OPTeprint = {https://aiche.onlinelibrary.wiley.com/doi/pdf/10.1002/aic.690050211},
year = {1959}
}

@article{Zampogna2016, title={Fluid flow over and through a regular bundle of rigid fibres}, volume={792}, DOI={10.1017/jfm.2016.66}, journal={Journal of Fluid Mechanics}, author={Zampogna, Giuseppe A. and Bottaro, Alessandro}, year={2016}, pages={5–35}}

@article{Box2020,
author = {Finn Box  and Gunnar G. Peng  and Draga Pihler-Puzović  and Anne Juel },
title = {Flow-induced choking of a compliant Hele-Shaw cell},
journal = {Proceedings of the National Academy of Sciences},
volume = {117},
number = {48},
pages = {30228-30233},
year = {2020},
doi = {10.1073/pnas.2008273117},
URL = {https://www.pnas.org/doi/abs/10.1073/pnas.2008273117},
OPTeprint = {https://www.pnas.org/doi/pdf/10.1073/pnas.2008273117}}

@article{Moore:2023,
  title = {Clogging of a Rectangular Slit by a Spherical Soft Particle},
  author = {Moore, Charles Paul and Husson, Julien and Boudaoud, Arezki and Amselem, Gabriel and Baroud, Charles N.},
  journal = {Phys. Rev. Lett.},
  volume = {130},
  issue = {6},
  pages = {064001},
  numpages = {5},
  year = {2023},
  month = {Feb},
  publisher = {American Physical Society},
  doi = {10.1103/PhysRevLett.130.064001},
  url = {https://link.aps.org/doi/10.1103/PhysRevLett.130.064001}
}

@inproceedings{Bergou:2008,
author = {Bergou, Mikl\'{o}s and Wardetzky, Max and Robinson, Stephen and Audoly, Basile and Grinspun, Eitan},
title = {Discrete elastic rods},
year = {2008},
isbn = {9781450301121},
publisher = {Association for Computing Machinery},
address = {New York, NY, USA},
url = {https://doi.org/10.1145/1399504.1360662},
doi = {10.1145/1399504.1360662},
booktitle = {ACM SIGGRAPH 2008 Papers},
articleno = {63},
number = {63},
numpages = {12},
location = {Los Angeles, California},
series = {SIGGRAPH '08}
}

\end{document}